\documentclass[article,amsmath,amssymb]{revtex4-2}
\usepackage{graphicx}
\usepackage{dcolumn}
\usepackage[colorlinks]{hyperref}
\usepackage{bm}

\begin{document}
\title{Cooperative decay of an ensemble of atoms in a one-dimensional chain with a single excitation}

\author{Nicola Piovella}
\affiliation{Dipartimento di Fisica "Aldo Pontremoli", Universit\`{a} degli Studi di Milano, Via Celoria 16, I-20133 Milano, Italy \&
INFN Sezione di Milano, Via Celoria 16, I-20133 Milano, Italy}

\begin{abstract}
We provide an analytic expression of the spectrum of the cooperative decay rate of $N$ two-level atoms regularly distributed on a ring in the single-excitation configuration. The results are obtained first for the scalar model and then extended to the vectorial light model, assuming all the dipoles are aligned.
\end{abstract}

\maketitle

\section{Introduction}

In this paper, we study the cooperative emission of an array of identical two-level atoms, organized in a circular ring of radius $\rho$. This study is part of the general subject of cooperative spontaneous emission by $N$ excited two-level atoms, extensively studied since the seminal work by Dicke in 1954 \cite{Dicke1954} and Lehmberg in 1970 \cite{Lehmberg1970}. It includes the well-known effect of superradiance, i.e., enhanced spontaneous emission  due to constructive interference between the emitters \cite{Bonifacio1971,Gross1982}, and subradiance, i.e., inhibited emission due to destructive interference between the emitters, which is more elusive and difficult to observe~\cite{Crubellier1985,Bienaime2012,Guerin2016,Ferioli2021,Cipris2021}. Subradiance  has seen a large increase in interest in the last few years, as it offers the opportunity of storing photons in emitter ensembles for times longer than the single emitter lifetime \cite{Scully2015,Jen2016,Facchinetti2016,Bettles2016,Asenjo2017,Needham2019,Reitz2022,Stiesdal2020}. In particular, superradiance and subradiance have been studied in the single-excitation configuration, belonging to the regime of linear optics.
In disordered systems, the cooperative decay must be studied numerically, usually by solving the dynamics of an initially excited ensemble \cite{Scully2006,Bienaime2013}. A more appealing situation is when the atoms form ordered arrays, where cooperativity may be enhanced.  For instance, infinite and finite linear chains of two-level atoms are considered in ref. \cite{Asenjo2017,Needham2019,Cech2023,Piovella2024}. {Subradiance has been experimentally observed also in a bidimensional lattice} \cite{Rui2020} or in an ensemble of atoms coupled to the guided mode of an optical nanofiber \cite{Pennetta2022}.

Having recently studied the finite chain of atoms \cite{Piovella2024}, we are interested here in studying the decay rates when the atoms form a closed configuration since it it expected to lead to a stronger suppression of the excitation. The cooperative single-quantum excitation of a closed-ring chain was studied in the past in ref.~\cite{Freedhoff1986} and more recently in \cite{Asenjo2017,Needham2019,Cech2023,Moreno2019}.
Starting from the effective non-Hermitian Hamiltonian, which includes an imaginary part describing the cooperative spontaneous decay and a real part describing the cooperative energy shift \cite{Lehmberg1970,Friedberg1973}, we focus on the cooperative decay only, calculating analytically the spectrum of the decay rates. The analysis is carried out by initially assuming the scalar model of light and neglecting the vectorial nature of the dipoles. The scalar model is particularly attractive because, since the polarization direction does not play any role, it is able to catch the main features of cooperativity just considering the relative phases of the emitters. Then, we extend the results to the vectorial light model for a set of $N$ equally oriented dipoles.

\section{Scalar Model}\label{s:model}

We consider $N$ identical two-level atoms with transition frequency $\omega_0=ck_0$, linewidth $\Gamma$, and dipole $\mu$. The atoms are prepared in a single-excitation state;  $|g_j\rangle$ and $|e_j\rangle$ are the ground and excited states, respectively, of the $j$-th atom, $j=1,\ldots,N$, which is placed at position $\mathbf{r}_j$.
We consider here the single-excitation effective Hamiltonian in the scalar approximation, whereas the exact vectorial model will be considered later. If we assume that only one photon is present, when tracing over the radiation degrees of freedom, the dynamics of the atomic system can be described by the non-Hermitian Hamiltonian \cite{Akkermans2008,Bienaime2013}
\begin{eqnarray}
\hat H & =&-i\frac{\hbar}{2}\sum_{j,m}G_{jm}\,
    \hat{\sigma}_j^\dagger\hat\sigma_m,\label{Heff}
\end{eqnarray}
 where $\hat\sigma_j=|g_j\rangle\langle e_j|$ and $\hat\sigma_j^\dagger=|e_j\rangle\langle g_j|$ are the lowering and raising operators, $G_{jm}$ is the scalar Green function,
\begin{equation}\label{gammajm}
    G_{jm}=
   	\left\{
    \begin{array}{ll}
    \Gamma_{jm}-i\, \Omega_{jm} & \mbox{if}~j\neq m, \\[1ex]
    \Gamma & \mbox{if}~j = m,
\end{array}
	\right.
\end{equation}
and
\begin{equation}\label{gammajm:bis}
    \Gamma_{jm} = \Gamma\frac{\sin(k_0r_{jm})}{k_0r_{jm}}\quad , \quad
    \Omega_{jm} = \Gamma\frac{\cos(k_0r_{jm})}{k_0r_{jm}},
\end{equation}
where $r_{jm}=|\mathbf{r}_j-\mathbf{r}_m|$.
$\hat H$ contains both real and imaginary parts, which takes into account that the excitation is not conserved since it can leave the system by emission. We focus our attention on the decay term $\Gamma_{jm}$.

\begin{figure}
\includegraphics[width=9 cm]{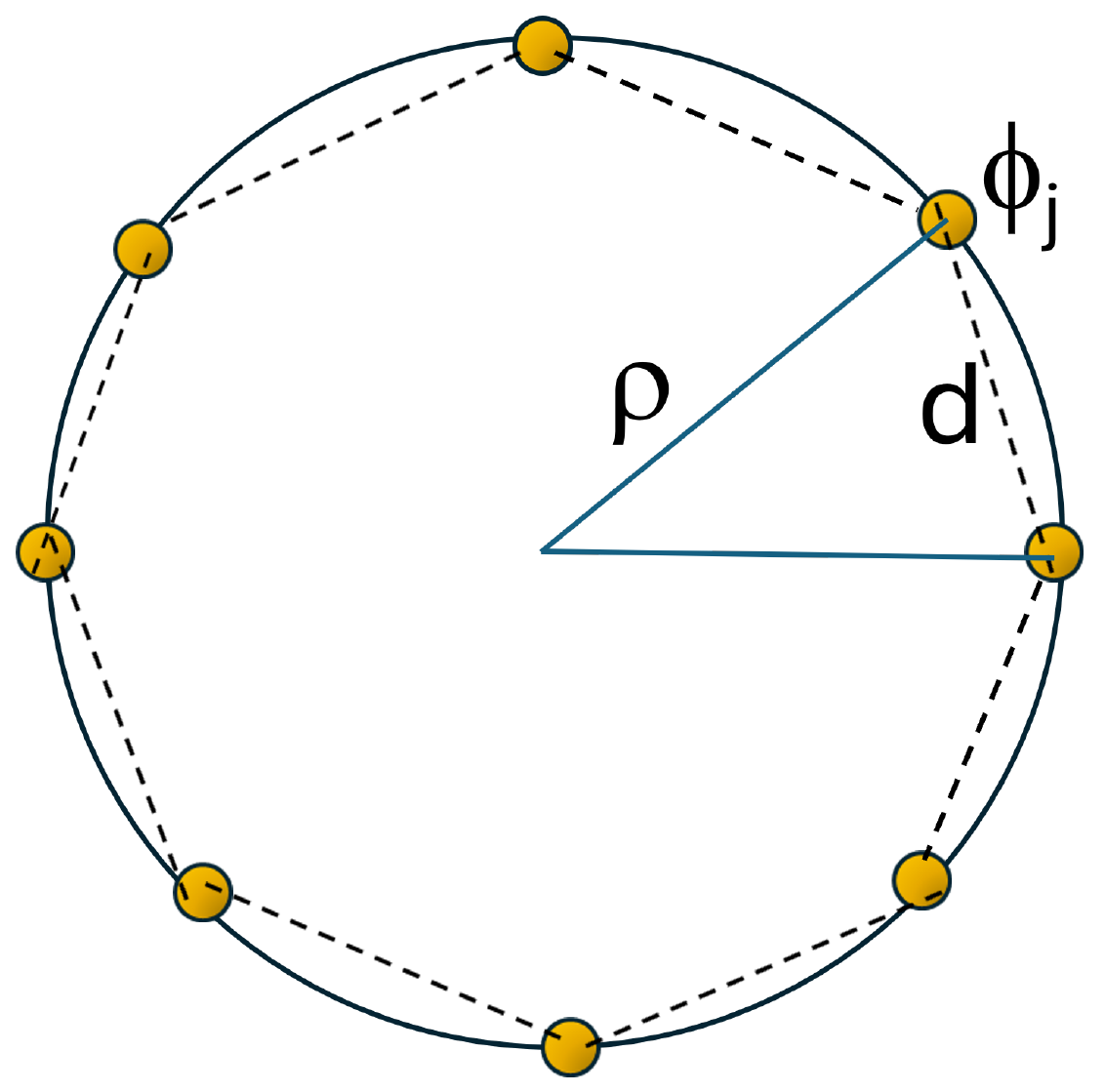}
        \caption{Scheme of the system: a ring with interparticle distance $d$, radius $\rho$, and angular position given by $\phi_j=2\pi(j-1)/N$}
        \label{setup}
\end{figure}

We consider $N$ atoms on a ring of radius $\rho$ and angles $\phi_j=(2\pi/N)(j-1)$, with $j=1,\dots,N$ (see Figure~\ref{setup}).
We write
\begin{equation}
\Gamma_{jm}=\Gamma\frac{\sin(k_0r_{jm})}{k_0r_{jm}}=\frac{\Gamma}{2}\int_{-1}^1 e^{i(k_0r_{jm})t}dt\label{int}
\end{equation}
where $k_0r_{jm}=2k_0\rho\sin(\phi_{jm}/2)$ and $\phi_{jm}=\phi_j-\phi_m$. Then, we expand the exponential
\begin{eqnarray*}
e^{2ia t\sin(\phi_{jm}/2)}&=&J_0(2a t)+2\sum_{n=1}^\infty J_{2n}(2at)\cos(n\phi_{jm})\\
&+&2i\sum_{n=1}^\infty J_{2n+1}(2a t)\cos[(2n+1)\phi_{jm}/2]
\end{eqnarray*}
where $a=k_0\rho$.
The last term is odd in $t$, so it can be dropped from the integral in Eq. (\ref{int}), obtaining
\begin{equation}
\Gamma_{jm}=\Gamma\sum_{n=-\infty}^{+\infty} c_n(a)e^{in(\phi_j-\phi_m)}\label{fact}
\end{equation}
where
\begin{equation}
c_n(a)=\int_0^1J_{2n}(2a t)dt=\frac{a^{2n}}{\Gamma(2+2n)}{}_1F_2[1/2+n;3/2+n,1+2n;-a^2]
\end{equation}
where ${}_1F_2[\alpha;\beta_1,\beta_2;z]$ is the Hypergeometric PFQ function and $\Gamma(n)$ is the gamma function.
Notice that $c_n(a)=c_{-n}(a)$ and
\begin{equation}
\sum_{n=-\infty}^{+\infty}c_n(a)=1
\end{equation}
so that $\Gamma_{jj}=\Gamma$. 
We define the spectrum of the decay rates as \cite{Moreno2019}
\begin{equation}
\Gamma_k=\frac{1}{N}\sum_{j,m=1}^N e^{ik\phi_{jm}}\Gamma_{jm}
\end{equation}
where $k=-N/2,\dots,N/2$ is discrete (let us suppose $N$ is even). Since by Eq. (\ref{fact}) the sum on $j$ and $m$ factorizes, we obtain
\begin{equation}
\Gamma_k=\frac{\Gamma}{N}\sum_{n=-\infty}^{+\infty}c_n(a)|F_{k+n}|^2
\end{equation}
where
\begin{equation}
|F_k|^2=\left|\sum_{j=1}^N e^{ik(2\pi/N)(j-1)}\right|^2=\frac{\sin^2(k\pi)}{\sin^2(k\pi/N)}=N^2\sum_{m=-\infty}^{+\infty}\delta_{k,mN}
\end{equation}
and
\begin{equation}
\Gamma_k=\Gamma N\sum_{m=-\infty}^{+\infty}c_{k-mN}(a)\label{Gamma:final}.
\end{equation}

The advantage of this expression is that is valid for an arbitrarily large number of atoms $N$. Furthermore, the spectrum is symmetric, with $\Gamma_{-k}=\Gamma_k$. Equation (\ref{Gamma:final}) is the main result of the paper.

\subsection*{Analysis}

\begin{figure}
\includegraphics[width=9 cm]{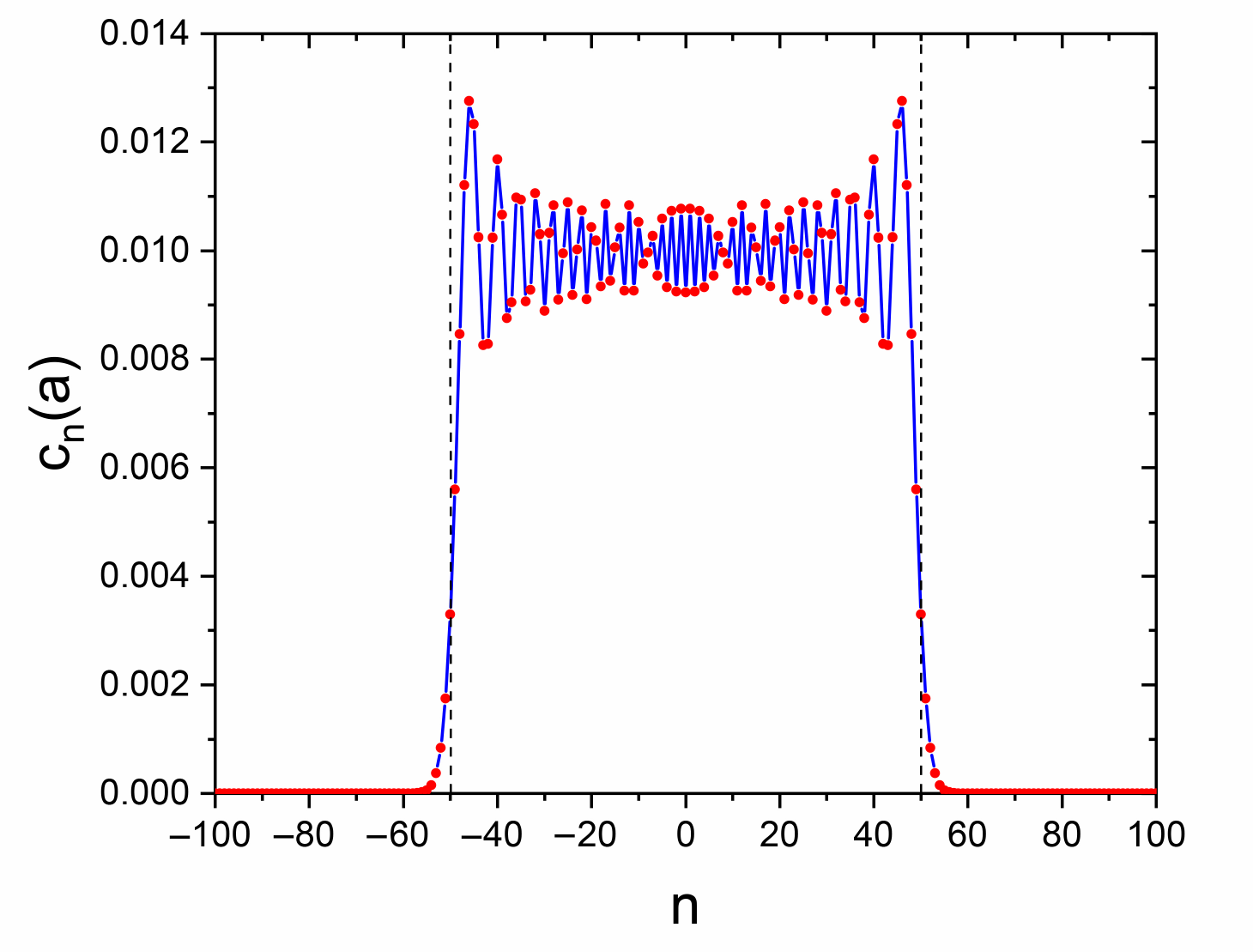}
        \caption{$c_n(a)$ vs. $n$ for $a=50$. Vertical dashed lines are for $n=\pm a$.}
        \label{fig1}
\end{figure}

It results that $c_n(a)\approx 0$ for $|n|>a$; for $a\rightarrow 0$, $c_n(a)=\delta_{n,0}$, whereas for $a\gg 1$, $c_n(a)\sim 1/(2a)$ for $|n|<a$ (see the example of Figure \ref{fig1} for $a=50$).

Hence, in the limit $a\rightarrow 0$, $\Gamma_k=\Gamma N \delta_{k,0}$: the spectrum is composed by a single superradiant component $k=0$ and the remaining subradiant components with $k\neq 0$  (Dicke limit). If $a<N/2$, $\Gamma_k\approx 0$ for $|k|>a$ (subradiance). Instead, assuming $a\gg N/2$, the sum of Equation~(\ref{Gamma:final}) spans from $-m_{\mathrm{max}}$ to $m_{\mathrm{max}}$, where $m_{\mathrm{max}}$ is determined by the condition $m_{\mathrm{max}}N-k\sim a$. Hence,
\begin{equation}
m_{\mathrm{max}}\sim \frac{a}{N}+\frac{k}{N}\sim \frac{a}{N}
\end{equation}
since $|k|<N/2$ and $a\gg N$. Then, Equation (\ref{Gamma:final}) gives, in the limit $a\gg N/2$,
\begin{equation}
\Gamma_k\sim \Gamma N\sum_{m=-m_{\mathrm{max}}}^{m_{\mathrm{max}}}c_{k-mN}(a)\sim\Gamma N \frac{1}{2a}\frac{2a}{N}\sim\Gamma.
\end{equation}
As expected,
 when $a/N$ becomes very large, the distance $d$ between adjacent atoms becomes much larger than the wavelength $\lambda_0$, and cooperativity disappears.
Ref. \cite{Asenjo2017} investigated numerically a ring distribution of $N$ atoms with fixed atom--atom separation $d$. In our case, the spectrum of the decay rate is a function of radius $\rho$ and of the number of atoms $N$. The distance $d$ between adjacent atoms is 
$d=2\rho\sin(\pi/N)\approx 2\pi\rho/N$, where the last expression is valid for large $N$. The previous necessary condition for subradiance, $k_0\rho< N/2$, implies $d/\lambda_0<0.5$. As discussed before, the sum in Equation~(\ref{Gamma:final}) spans from $-m_{\mathrm{max}}$ to $+m_{\mathrm{max}}$, where $m_{\mathrm{max}}$ is determined by $m_{\mathrm{max}}N-k\sim k_0\rho\sim (d/\lambda_0)N$ so that $m_{\mathrm{max}}\sim (k/N)+(d/\lambda_0)$. Taking the most subradiant value $k=N/2$ (since subradiance occurs for $(d/\lambda_0)N<k<N/2$), then $m_{\mathrm{max}}\sim (d/\lambda_0)+0.5<1$, and the only term surviving in the sum of Equation (\ref{Gamma:final}) is $m=0$:
\begin{equation}
\Gamma_{N/2}\sim\Gamma Nc_{N/2}(dN/\lambda_0)\label{Gamma:sub:1}
\end{equation}

For $N\gg 1$,
\begin{equation}
\Gamma_{N/2}\sim\Gamma J_{N}(2dN/\lambda_0)\sim\frac{1}{\sqrt{2\pi N}}(e d/\lambda_0)^N,
\end{equation}
where $J_n(x)$ is the Bessel function of order $n$, and the last expression is valid when $ed/\lambda_0<1$. In conclusion, the spontaneous emission is exponentially suppressed, as seen in ref. \cite{Asenjo2017}, if we increase $N$, keeping the distance between adjacent atoms constant, i.e., increasing the radius $\rho$ proportionally to $N$.
On the contrary, in the continuous limit, we tend to let $N\rightarrow\infty$ and $d/\lambda_0\rightarrow 0$ such that $k_0\rho$ remains finite. Then, in the sum of Equation (\ref{Gamma:final}), there remains only the term $m=0$, and
\begin{equation}
\Gamma_k^{\mathrm{cont}}\sim\Gamma Nc_{k}(k_0\rho)\label{Gamma:sub:2}
\end{equation}

Subradiance occurs for $k_0\rho<|k|<N/2$, whereas  superradiance occurs for $|k|<k_0\rho$. 
When $k_0\rho\rightarrow 0$, then $\Gamma_k^{\mathrm{cont}}\sim\Gamma N\delta_{k,0}$, whereas when $k_0\rho\gg 1$, then $\Gamma_k^{\mathrm{cont}}\sim\Gamma N/(2k_0\rho)$ when $|k|<k_0\rho$.

\begin{figure}
\includegraphics[width=10 cm]{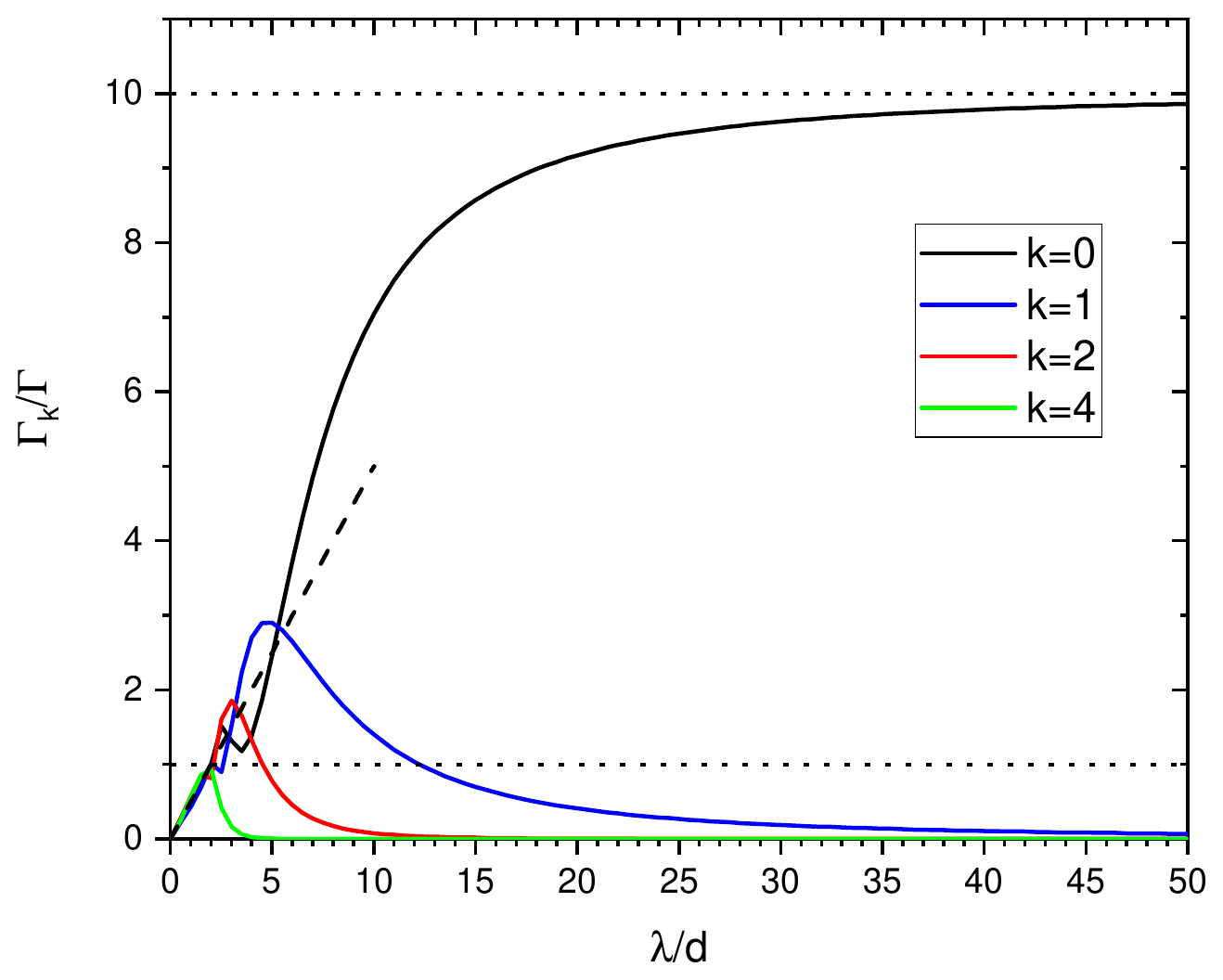}
        \caption{$\Gamma_k/\Gamma$ vs $\lambda_0/d$ for $N=10$ and $k=0,1,2,4$. In the Dicke limit, $\lambda_0/d\rightarrow\infty$, only the superradiant $k=0$ with a decay rate $N\Gamma$ is present, and $N-1$ modes are completely subradiant. In the limit $\lambda_0/d\rightarrow 0$, all the modes decay with the rate $\Gamma_k/\Gamma=\lambda_0/2d$ (dashed line).}
        \label{fig2}
\end{figure}

In Figure \ref{fig2}, we plot the decay rate for different values of $k=0,1,2,4$ for $N=10$ as a function of $\lambda_0/d$. We see that as $\lambda_0/d\rightarrow 0$, $\Gamma_k\approx \Gamma N/2a\sim (\Gamma/2)(\lambda_0/d)$ (dashed line), whereas for $\lambda_0/d\rightarrow\infty$, the emitters are so close that the range of interaction is effectively infinite, yielding a single superradiant mode decaying at rate $N\Gamma$ and $N-1$ perfectly subradiant modes. This may be explained since $\Gamma_k\approx 0$ for $k>a$; since $a\approx N(d/\lambda_0)$, when $\lambda_0/d\rightarrow\infty$, $a\rightarrow 0$ and the mode $k=0$ decays superradiantly as $N\Gamma$, whereas the other modes with $k>0$ are dark. Conversely, when $\lambda_0/d\ll  1$, $a\gg 1$ and $\Gamma_k\sim N\Gamma/(2a)\sim \Gamma(\lambda_0/2d)$. The results are in agreement with those presented in Ref. \cite{Moreno2019}.

\section{Vectorial Model}

We now extend the previous expressions to the vectorial model, taking into account the polarization of the electromagnetic field. The non-Hermitian Hamiltonian is now
\begin{eqnarray}
\hat H & =&-i\frac{\hbar}{2}\sum_{\alpha,\beta}\sum_{j,j'}G_{\alpha,\beta}(\mathbf{r}_j-\mathbf{r}_{j'})\,
    \hat{\sigma}_{j,\alpha}^\dagger\hat\sigma_{j',\beta}.\label{H_vec}
\end{eqnarray}
where $\alpha,\beta=(x,y,z)$. Here  $\hat\sigma_{j,x}=(\hat\sigma_{j}^{m_J=1}+\hat\sigma_{j}^{m_J=-1})/2$, $\hat\sigma_{j,y}=(\hat\sigma_{j}^{m_J=1}-\hat\sigma_{j}^{m_J=-1})/2i$ and $\hat\sigma_{j,z}=\hat\sigma_{j}^{m_J=0}$, where 
 $\hat\sigma_{j}^{m_J}=|g_j\rangle\langle e_{j}^{m_J}|$ is the lowering operator between the ground state $|g_j\rangle$ and the three excited states $|e_{j}^{m_J}\rangle$ of the $j$th atom with quantum numbers $J=1$ and $m_J=(-1,0,1)$. The vectorial Green function in Eq. (\ref{H_vec}) is 
\begin{equation}
G_{\alpha,\beta}(\mathbf{r})=\frac{3\Gamma}{2}\frac{e^{ik_0r}}{ik_0r}\left[\delta_{\alpha,\beta}-\hat n_\alpha\hat n_\beta
+\left(\delta_{\alpha,\beta}-3\hat n_\alpha\hat n_\beta\right)\left(\frac{i}{k_0 r}-\frac{1}{k_0^2 r^2}\right)
\right]
\end{equation}
with $r=|\mathbf{r}|$ and ${\hat n}_\alpha$ being the components of the unit vector $\hat{\mathbf{n}}=\mathbf{r}/r$. 
We consider a ring with $r_{jm}=2\rho\sin(\phi_{jm}/2)$ and all the dipoles aligned with an angle $\delta$ with respect to ring's plane so that $\hat n_\alpha=\hat n_\beta=\cos\delta$ and
\begin{equation}
G^{(\delta)}(r_{jm})=\frac{3\Gamma}{2}\frac{e^{ik_0r_{jm}}}{ik_0r_{jm}}\left[\sin^2\delta
+(1-3\cos^2\delta)\left(\frac{i}{k_0r_{jm}}-\frac{1}{k_0^2r_{jm}^2}\right)\right].\label{Gaa}
\end{equation}

The decay rate for the vectorial model is given by the real part of $G^{(\delta)}(r_{jm})$,
\begin{eqnarray}
\Gamma^{(\delta)}(r_{jm})&=&
\frac{3\Gamma}{2}\left[\sin^2\delta j_0(k_0r_{jm})+(3\cos^2\delta-1)\frac{j_1(k_0r_{jm})}{k_0r_{jm}}\right]
\end{eqnarray}
where $j_0(x)=\sin x/x$ and $j_1(x)=\sin x/x^2-\cos x/x$ are the spherical Bessel functions of the orders $n=0$ and $n=1$.
By using the identities
\begin{eqnarray*}
j_0(x) &=&\frac{1}{2}\int_{-1}^{1}e^{ixt}dt\\
j_0(x)-2\frac{j_1(x)}{x}&=&\frac{1}{2}\int_{-1}^1 t^2 e^{ixt}dt,
\end{eqnarray*}
we can write
\begin{eqnarray}
\Gamma^{(\delta)}(r_{jm})&=&\frac{3\Gamma}{8}\left\{(1+\cos^2\delta)\int_{-1}^1 e^{i(k_0r_{jm})t}dt +(1-3\cos^2\delta)
\int_{-1}^1 t^2e^{i(k_0r_{jm})t}dt
\right\}\nonumber\\
&=&\frac{3\Gamma}{4}\sum_{n=-\infty}^\infty \left\{(1+\cos^2\delta)c_n(a)+(1-3\cos^2\delta)
d_n(a)\right\}e^{in(\phi_j-\phi_m)}
\end{eqnarray}
where $a=k_0\rho$ and
\begin{equation}
d_n(a)=\int_0^1 t^2J_{2n}(2a t)dt=
\frac{a^{2n}}{2}\frac{\Gamma(3/2+n)}{\Gamma(1+2n)\Gamma(5/2+n)}{}_1F_2[3/2+n;5/2+n,1+2n;-a^2]\label{dn}
\end{equation}

As before, the spectrum is
\begin{equation}
\Gamma_k^{(\delta)}=\frac{1}{N}\sum_{j,m=1}^N e^{ik\phi_{jm}}\Gamma_{jm}^{(\delta)}
\end{equation}
where $k=-N/2,\dots,N/2$ is discrete (let suppose $N$ even). Then,
\begin{equation}
\Gamma_k^{(\delta)}=\frac{3\Gamma}{4N}\sum_{n=-\infty}^{+\infty}
\left\{(1+\cos^2\delta)c_n(a)+(1-3\cos^2\delta)
d_n(a)\right\}
|F_{k+n}|^2
\end{equation}
and
\begin{equation}
\Gamma_k^{(\delta)}=\frac{3\Gamma N}{4}\sum_{m=-\infty}^{+\infty}
\left\{(1+\cos^2\delta)c_{k-mN}(a)+(1-3\cos^2\delta)
d_{k-mN}(a)\right\}
\end{equation}

\begin{figure}
\includegraphics[width=9 cm]{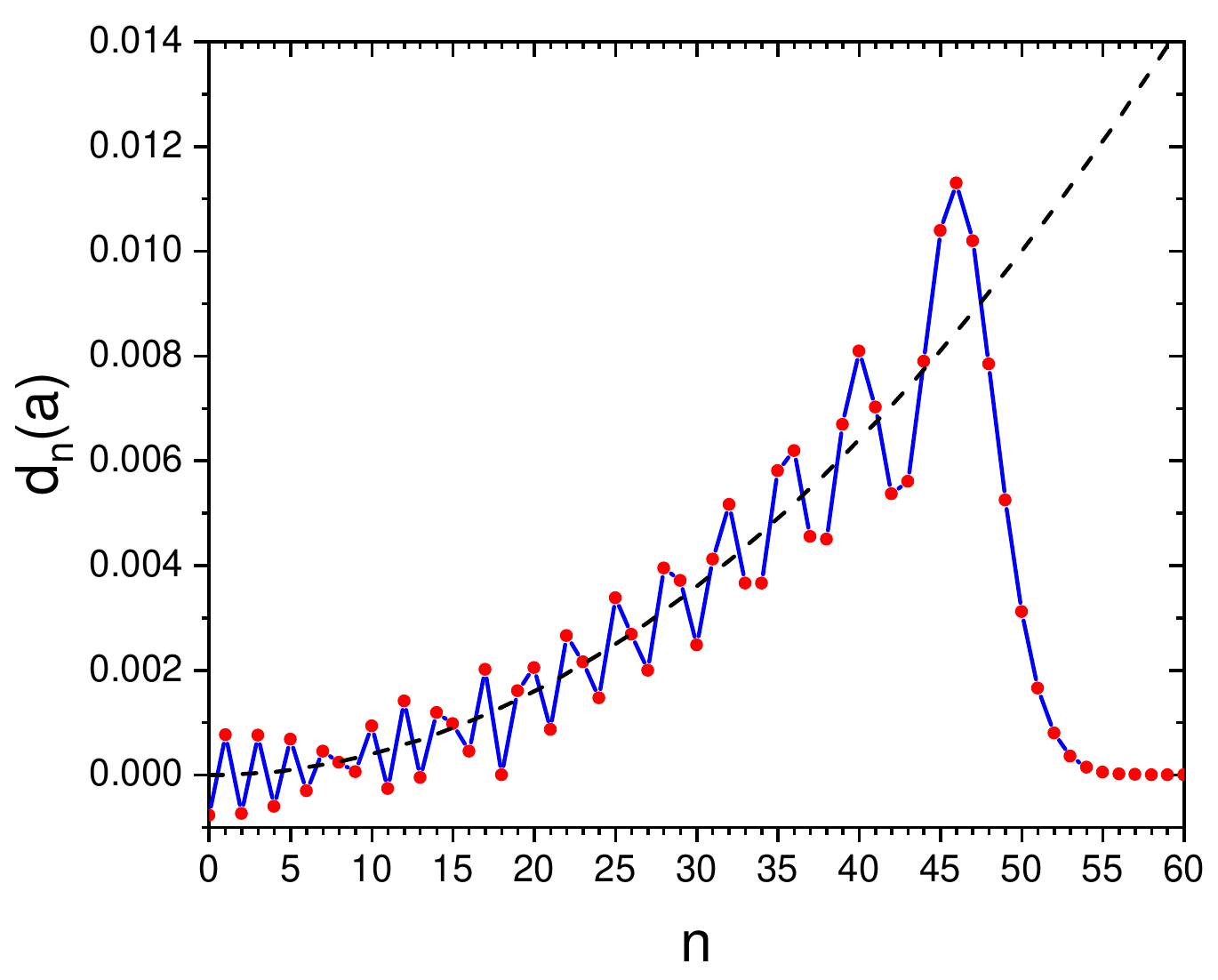}
        \caption{$d_n(a)$ vs. $n$ for $a=50$. The red dots are the exact solution of Eq.(\ref{dn}) and the dashed line is the approximated solution $d_n(a)\sim 1/(2a^3)(n^2-1/4)$.}
        \label{fig3}
\end{figure}

It results again  that $d_n(a)\approx 0$ for $|n|>a$;  
for $a\rightarrow 0$, $d_n(a)=(1/3)\delta_{n,0}$, whereas for $a\gg 1$, $d_n(a)\sim 1/(2a^3)(n^2-1/4)$ for $|n|<a$ (see the example of Figure \ref{fig3} for $a=50$).

Hence, in the limit $a\rightarrow 0$, $\Gamma_k^{(\delta)}=\Gamma N$, whereas for $a\gg 1$ and $|k|<a$,
\begin{equation}
\Gamma_k^{(\delta)}=\frac{3\Gamma N}{8a}
\left\{1+\cos^2\delta+\frac{1}{a^2}(1-3\cos^2\delta)(k^2-1/4)\right\}
\end{equation}

\begin{figure}
\includegraphics[width=9 cm]{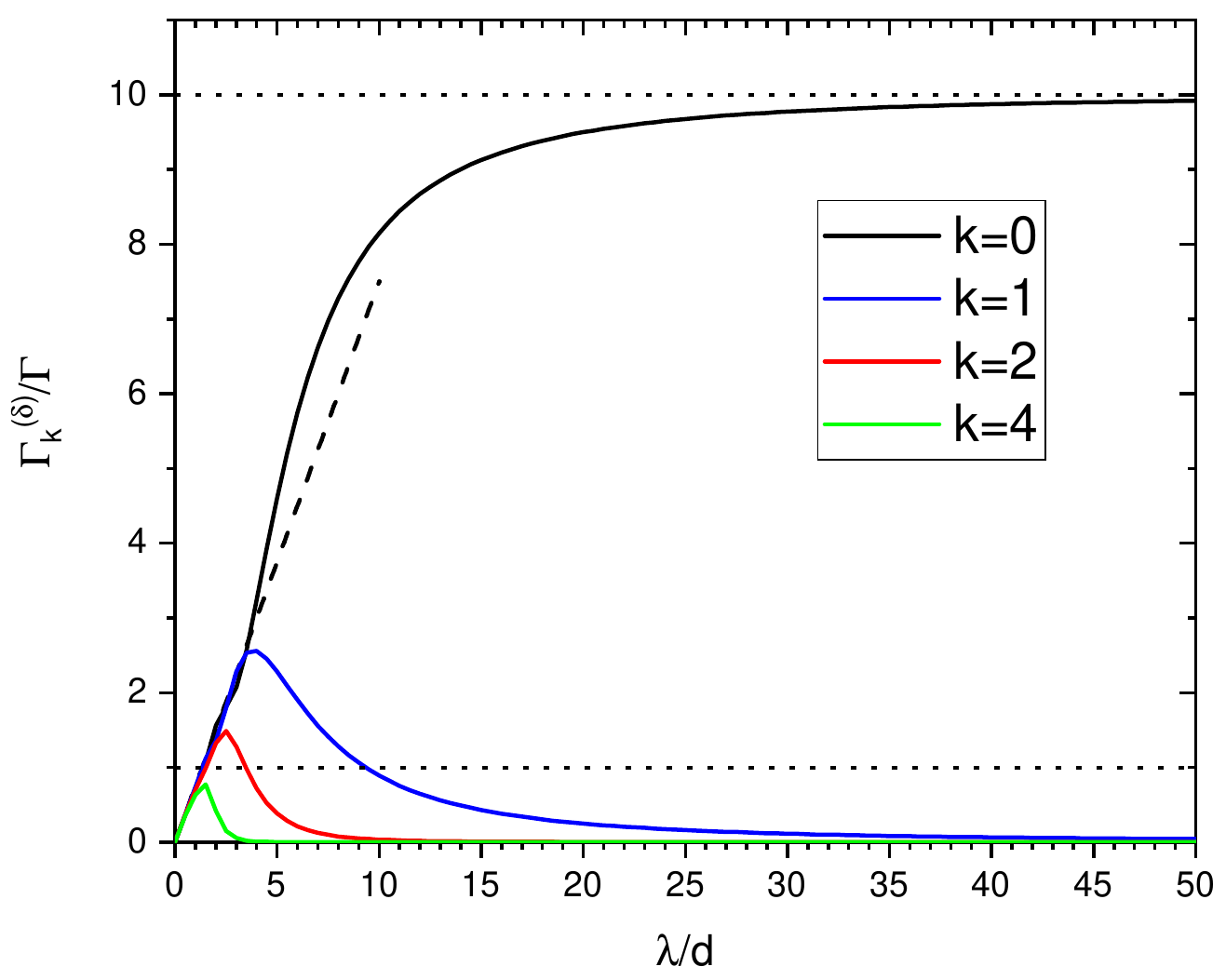}
        \caption{$\Gamma_k^{(\delta)}/\Gamma$ vs. $\lambda_0/d$ for $N=10$ and $k=0,1,2,4$, for $\delta=0$. In the Dicke limit, $\lambda_0/d\rightarrow\infty$, only the superradiant $k=0$ with a decay rate $N\Gamma$ is present, and $N-1$ modes are completely subradiant. In the limit $\lambda_0/d\rightarrow 0$, all the modes decay with the rate $\Gamma_k^{(\delta)}/\Gamma=(3/4)(\lambda_0/d)$ (dashed line).}
        \label{fig4}
\end{figure}

As for the scalar model, we  plot in Figure~\ref{fig4} the decay rate $\Gamma_k^{(\delta)}$ (in units of $\Gamma$) for $N=10$ and for different values of $k=0,1,2,4$ as a function of $\lambda_0/d$ for $\delta=0$. The curves are similar to those of the scalar model. The only difference is that as $\lambda_0/d\rightarrow 0$, $\Gamma_k^{(\delta)}\approx (3\Gamma/4)(\lambda_0/d)$ (dashed line in Figure~\ref{fig4}).

\unskip

\section{Conclusions}

In conclusion, we have presented an analytical expression for the discrete spectrum $\Gamma_k$ of the decay rates of $N$ atoms regularly distributed on a ring in the single-excitation configuration, both in the scalar and in the vectorial model. The results are in agreements with those presented in Ref. \cite{Asenjo2017,Moreno2019}, where the rates are calculated numerically as the imaginary part of the eigenvalues of the effective Hamiltonian.
The analytical expression shows that the decay rates are proportional to $N$ times a function of the parameter $a=k_0\rho$, where $\rho$ is the ring's radius. The modes with $|k|>a$ are dark, with almost zero decay rates, whereas the modes with $|k|<a$ are superradiant, with $\Gamma_k\sim N\Gamma/(2a)$ for large values of $a$. Keeping fixed the distance $d\sim 2\pi\rho/N$ between adjacent atoms on the ring, the most subradiant modes decay exponentially with $N$ when $d/\lambda_0<0.5$ as seen in Ref. \cite{Asenjo2017}. Is it interesting to note that in the ring configuration, the subradiant spectrum depends on the parameter $a$ and not on the atomic number $N$, as it occurs, for instance, in a finite linear chain \cite{Piovella2024}.

{These analytical results may be important for the future implementation in experimental setups. While ring configurations are not so easy to implement using individual atoms in optical traps, closely related ring-shaped structures of dipoles appear naturally in biological light harvesting complexes} \cite{Law2004} {or can be set up using quantum dot micro-arrays}~\cite{Richner2016}. {Alternatively, one could study such structures in tweezer arrays}  \cite{Barredo2016}.
{As a potential utility of the present results, it has been shown that subradiance may reduce errors for applications such as quantum memories, enabling more efficient quantum information processing and quantum communication protocols} \cite{Manzoni2018}, {or may greatly extend the excited-state lifetimes in optical-lattice clocks} \cite{Henriet2019}.

\end{document}